\documentclass[aps,prd, floatfix,twocolumn,preprintnumbers,superscriptaddress,showpacs]{revtex4-1}

\usepackage{amsmath,amsthm,amsfonts,slashed,setspace}
\usepackage{epsfig,bbold,latexsym,yfonts,float}
\usepackage{color}
\usepackage[shortlabels]{enumitem}
\usepackage{gensymb,verbatim}
\usepackage{cancel}
\usepackage{multirow}
\usepackage{xcolor}
\usepackage{amssymb}
\usepackage{mathrsfs} 
\makeatletter
\def\mathcolor#1#{\@mathcolor{#1}}
\def\@mathcolor#1#2#3{%
  \protect\leavevmode
  \begingroup
    \color#1{#2}#3%
  \endgroup
}
\makeatother

\newcommand{\euv}{\varepsilon_{\rm UV}}

\usepackage{physics}

\newcommand{\nn}{\nonumber    \\ }

\begin{document}


\author{Ian Balitsky, Wayne Morris and  Anatoly  Radyushkin}
\address{Physics Department, Old Dominion University, Norfolk,
             VA 23529, USA}
\address{Thomas Jefferson National Accelerator Facility,
              Newport News, VA 23606, USA
}

\title{Short-Distance Structure of Unpolarized Gluon   Pseudodistributions}

\begin{abstract}

We present the results that form the basis for 
calculations of the  unpolarized gluon parton distributions (PDFs)  using the pseudo-PDF approach.
We  give the  results  for the most complicated box diagram both 
 for  gluon bilocal operators with   arbitrary indices,
and  for combinations of indices corresponding to three matrix elements
that are most convenient to extract the twist-2 invariant amplitude.  
We also present detailed results for the gluon-quark transition diagram.
The additional results for the box diagram and the gluon-quark  contribution 
 may be used for extractions of the gluon PDF from  different  matrix elements,
with a possible cross-check of the results obtained in this way.

              
\end{abstract}

\maketitle


\section{
 Introduction}

Extraction   of  the parton distribution functions (PDFs) from  lattice calculations  attracts  now a  considerable interest
(see Refs.  \cite{Constantinou:2020hdm,Cichy:2018mum}  for  recent reviews and references to extensive literature).
Modern  efforts  aim at getting  PDFs $f(x)$  themselves rather than 
their $x^N$ moments. The recent progress in this endeavor  has been  stimulated 
by 
 the paper  \cite{Ji:2013dva} of  X. Ji. Its basic   
  proposal is  to consider equal-time versions of nonlocal operators 
 that define parton functions, such as  PDFs, distribution amplitudes, generalized  parton distributions, and 
 transverse momentum dependent distributions. 
In the case of ordinary PDFs,  
the  major object 
 of Ji's approach 
 is  a    ``parton quasi-distribution''  (quasi-PDF) $Q(y,p_3)$  \cite{Ji:2013dva,Ji:2014gla}. 
They produce 
PDFs  obtained in  the  large-momentum $p_3 \to \infty$  limit 
of quasi-PDFs. 

Alternatively, one may use coordinate-space oriented methods,   namely,  
 the ``good lattice cross sections'' approach
 \cite{Ma:2014jla,Ma:2017pxb}, the  Ioffe-time  analysis of equal-time correlators  \cite{Braun:2007wv,Bali:2017gfr,Bali:2018spj}
and  the pseudo-PDF approach \cite{Radyushkin:2017cyf,Radyushkin:2017sfi,Orginos:2017kos}. 
In these cases, parton distributions  are extracted through  taking  the 
short-distance $z_3 \to 0$ limit.

 In converting    the Euclidean  lattice data into
the    light-cone PDFs one should take into account that 
both the $p_3\to \infty$ and $z_3\to 0$ limits are  singular, and 
one needs to incorporate  {\it matching relations} 
to perform the conversion.

 The matching conditions in the quasi-PDF approach,  were studied 
for quark    \mbox{\cite{Ji:2013dva,Xiong:2013bka,Ji:2015jwa,Izubuchi:2018srq}} and gluon 
 PDFs \cite{Wang:2017eel,Wang:2017qyg,Wang:2019tgg}, for  the
 pion distribution amplitude (DA)  \cite{Ji:2015qla} and generalized parton distributions (GPDs) 
\cite{Ji:2015qla,Xiong:2015nua,Liu:2019urm}.

 The  matching relations in pseudo-PDF approach 
  were also derived in several cases, in particular,  for  non-singlet 
PDFs   \cite{Ji:2017rah,Radyushkin:2017lvu,Radyushkin:2018cvn,Zhang:2018ggy,Izubuchi:2018srq}.
The procedure of  lattice extraction  of  non-singlet GPDs and the pion DA
within  the pseudo-PDF framework  was outlined 
in  Ref.   \cite{Radyushkin:2019owq},
where the  relevant matching conditions  have been also derived. 

In our earlier paper \cite{Balitsky:2019krf} (see also Ref. \cite{Balitsky:2021bds}) 
we have outlined  the basic points  of pseudo-PDF  approach to extraction of unpolarized gluon PDFs, 
and have presented the one-loop  matching conditions for a particular  combination of gluon matrix 
elements,  that has the ``cleanest'' projection on the twist-2 gluon PDF. 
These results have been used already 
in lattice extractions of the unpolarized gluon PDFs 
in Refs.  \cite{Fan:2020cpa,Fan:2021bcr}  and  \cite{HadStruc:2021wmh}.

However, because of the letter nature of Ref.  \cite{Balitsky:2019krf}, 
we have skipped there some intermediate expressions and also results for two other matrix elements
that may be used for the gluon PDF extraction. 

In the present paper,  we  present a full result  for the most lengthy contribution of the ``box''  diagram,
and also its projections onto all 3 matrix elements containing the ``twist-2'' invariant amplitude.
We also give more details  about our calculations of the gluon-quark mixing corrections both 
for these matrix elements and for matrix elements with arbitrary indices.
The additional results  given in the present
paper may be used for extractions of the gluon PDF from two other matrix elements,
which may  give a 
 possibility to  cross-check  the results obtained from different matrix elements.

  The paper is organized as follows. 
In 
 \mbox{Section  II},   we  study the structure  of the matrix  elements of the gluonic bilocal operators.
 In particular, we identify 
 those that contain information about the twist-2 gluon PDF.
   In \mbox{Section III,}  we  discuss one-loop corrections,  and   specific properties of their  ultraviolet  and short-distance behavior.
 In subsection IIIf and Appendix A, we present our   results for the most complicated  ``box'' diagram.  
 The subject of Section IV  is the structure  of  perturbative evolution 
 of the gluon operators, gluon-quark mixing  and matching conditions.
 The result for the gluon-quark contribution generated by the gluon bilocal  operator with arbitrary indices 
 is given in Appendix B.
 Section V   contains    summary of the paper.

\setcounter{equation}{0}


\section{Matrix elements}

We are going to consider the  nucleon spin-averaged  matrix elements for operators composed of  two-gluon-fields   in the most general case when all
 four indices  are not contracted     
    \begin{align}
 { M}_{\mu \alpha;  \nu \beta }  (z,p) \equiv \langle  p |  \,
 G_{\mu \alpha} (z)  \, [z,0]\,  G_{ \nu \beta } (0) | p \rangle \  , 
 \label{ME}
\end{align}
 where  $
[z,0]$ stands for usual   straight-line gauge link 
 in the gluon  (adjoint) 
 representation
 \begin{align}
[x,y]~\equiv~{\rm Pexp}\Big\{ig\!\int_0^1\! dt~(x-y)^\mu \tilde{A}_\mu(tx+(1-t)y)\Big\}
 \  . 
 \label{straightE}
\end{align}

\subsection{Invariant amplitudes}

 We want to decompose ${ M}_{\mu \alpha;  \nu \beta }  (z,p) $  in several  tensor structures accompanied  by corresponding  invariant amplitudes.
 The  latter     
 may be built from two  available 4-vectors, namely  $p_\alpha$, $z_\alpha$,   and the 
 metric tensor $g_{\alpha \beta}$. 
 Building the tensors, we  incorporate  the antisymmetry   of $G_{\rho \sigma}$ with  respect to its indices. 
 This gives  \cite{Balitsky:2019krf}
    \begin{align}
& { M}_{\mu \alpha;   \nu  \beta}  (z,p)= \nonumber \\  & 
 \left( g_{\mu\nu} p_\alpha p_\beta - g_{\mu\beta} p_\alpha p_\nu - g_{\alpha\nu} p_\mu p_\beta + g_{\alpha\beta} p_\mu p_\nu \right) \mathcal{M}_{pp}  \nonumber \\
+& \left( g_{\mu\nu} z_\alpha z_\beta - g_{\mu\beta} z_\alpha z_\nu - g_{\alpha\nu} z_\mu z_\beta + g_{\alpha\beta} z_\mu z_\nu \right) \mathcal{M}_{zz}     \nonumber \\
+ & \left( g_{\mu\nu} z_\alpha p_\beta - g_{\mu\beta} z_\alpha p_\nu - g_{\alpha\nu} z_\mu p_\beta + g_{\alpha\beta} z_\mu p_\nu \right) \mathcal{M}_{zp}    \nonumber \\
+ & \left( g_{\mu\nu} p_\alpha z_\beta - g_{\mu\beta} p_\alpha z_\nu - g_{\alpha\nu} p_\mu z_\beta + g_{\alpha\beta} p_\mu z_\nu \right) \mathcal{M}_{pz}     \nonumber \\
+&   \left( p_\mu z_\alpha  - p_\alpha z_\mu\right) \left(  p_\nu z_\beta -  p_\beta z_\nu\right) \mathcal{M}_{ppzz} 
\nn  + &   \left(g_{\mu\nu} g_{\alpha\beta} -g_{\mu\beta} g_{\alpha\nu} \right)\mathcal{M}_{gg}    \ .
\label{Manb}
\end{align}
The  amplitudes ${\cal M}$ are functions of 
the Lorentz invariants of the problem, i.e. the  invariant interval $z^2$ and the Ioffe time \cite{Ioffe:1969kf} 
$(pz)\equiv - \nu$   (for further  convenience we define $\nu$ with the minus sign).

Since  the matrix element should be symmetric with respect to interchange of the
fields,  
the  functions $\mathcal{M}_{pp}$,  $\mathcal{M}_{zz}$, $\mathcal{M}_{gg} $,  $\mathcal{M}_{ppzz} $ and 
$ \mathcal{M}_{pz} - \mathcal{M}_{zp}  $  are
even functions of $\nu$, while   $ \mathcal{M}_{pz} + \mathcal{M}_{zp}  $ is
odd in $\nu$.  

\subsection{``Twist-2'' projection}

The  standard  light-cone gluon distribution $f_g(x)$ is defined through the convolution 
 $ g^{\alpha \beta} { M}_{+ \alpha;  \beta  +}  (z,p) $,  with the separation $z$ taken in the light-cone ``minus'' direction,
 $z=z_-$:
   \begin{align}
 g^{\alpha \beta} { M}_{+ \alpha;  \beta +}  (z_-,p) =  p_+ ^2  \int_{-1}^1 \dd x \, e^{i x p_+ z_-} x f_g (x)  \ . 
 \label{fgdef}
\end{align}

 Extracting  the  projection $ g^{\alpha \beta} { M}_{+ \alpha;  \beta +} $ from the decomposition (\ref{Manb}), we get 
  \begin{align}
 g^{\alpha \beta} { M}_{+ \alpha;  \beta +}  (z_-,p) =-2 p_+ ^2  \mathcal{M}_{pp}(\nu ,0)  \ . 
 \label{lcpp}
\end{align}
This means that   the gluon  PDF  is  determined by the ${\cal M}_{pp} $ invariant amplitude
  \begin{align}
- \mathcal{M}_{pp} (\nu,0) =\frac12
 \int_{-1}^1 \dd x \, e^{-i x \nu} x f_g (x) \,  \ . 
 \label{glPDF}
\end{align}

In view of Eq. (\ref{glPDF}), our   strategy is to  choose matrix elements  ${ M}_{\mu \alpha;   \lambda  \beta} $ 
that  contain
 ${\cal M}_{pp} $
in its  parametrization, and ideally nothing else.

Having in mind  lattice calculations, it is convenient to split 
the ``+'' components  
onto sum of  space- and time-components.  Also, 
due to  antisymmetry of $G_{\rho \sigma}$ with  respect to its indices,
 the combination  
$g^{\alpha \beta}  { M}_{+ \alpha; \beta +}  (z,p)$ includes  summation over the 
transverse indices $i,j =1,2$ only,  and reduces 
to 
  \begin{align}
&g^{ij}  { M}_{+ i;  j +} = - { M}_{+ 1;  1 +} -  { M}_{+ 2;  2 +}  \nn & =   { M}_{0 i;  0i}  +    { M}_{3 i;   3i}  +   ({ M}_{0 i;  3i}  +
 { M}_{3 i;    0i}  ) \ . 
\label{ii}
\end{align}
with summation over $i=1,2$ implied.

\subsection{Picking out $ \mathcal{M}_{pp}$ amplitude} 

As found in Ref.   \cite{Balitsky:2019krf}, there is an extension of the  $ { M}_{0 i;  i 0 }$ matrix element 
 that contains the $ \mathcal{M}_{pp}$ amplitude only,  
 \begin{align}
 { M}_{0 i;  i 0 }  +  { M}_{j i ;  i j } =  & 2   p_0^2  \mathcal{M}_{pp}   \  , 
 \label{00m}
\end{align}
 where the   summation  both over $i$ and $j$ is implied.
 
 One can apply a similar procedure on ${ M}_{3i ; i3}$.  Using the expression 
\begin{align}
{ M}_{30;03}&  \equiv  \bra{p} G_{30} (z) G_{03}(0) \ket{p}  \nn &=  m^2 \mathcal{M}_{pp}  -  z_3^2  \mathcal{M}_{zz}  -  z_3 p_3 \left( \mathcal{M}_{zp} +   \mathcal{M}_{pz}\right)
\nn &  -   p_0^2 z_3^2 \mathcal{M}_{ppzz}  + \mathcal{M}_{gg}  \ , 
\end{align}
we  construct  the combination:
\begin{align}
 { M}_{3 i;  i 3 }  +  2{ M}_{30 ;  03} = & 2   p_0^2  \mathcal{M}_{pp} -  2 p_0^2 z_3^2 \mathcal{M}_{ppzz} \ ,
\label{33m}
\end{align}
which  still has an additional term 
 proportional to the $\mathcal{M}_{ppzz}$ invariant amplitude. 
 Another minimally contaminated combination is given by 
\begin{align}
{ M}_{0 i;   i 3 } +  &{ M}_{3 i;  i 0 }  =  4 p_0 p_3  \mathcal{M}_{pp} +  2 p_0   z_3 
 \left(\mathcal{M}_{pz}+   \mathcal{M}_{zp}  \right)  \  . 
 \label{03m}
\end{align}

\subsection{Multiplicatively renormalizable combinations}

Off the light cone, the $ { M}_{\mu \alpha;   \lambda  \beta}$  matrix elements have extra  ultraviolet divergences
related to presence of the gauge link.
 For any   set of its indices $\{ \mu \alpha;  \lambda \beta  \}$, each matrix element 
 is multiplicatively renormalizable with respect to these divergences \cite{Li:2018tpe}, but 
 in general,  with different 
 anomalous dimensions.

In Ref. \cite{Zhang:2018diq}, it was established that  the combinations  represented in Eq. (\ref{ii}),  namely,  
$ { M}_{0 i; i 0 }  $,    $ { M}_{3 i;  i 3 } $,    \mbox{${ M}_{0 i; i3 }  +
 { M}_{3 i; i 0 }  $} (and also  \mbox{${ M}_{0 i; i 3}  -
 { M}_{3 i; i 0 }  $}),  with   
 summation over     transverse indices $i$,  
 are each  multiplicatively renormalizable at the one-loop level. 
 Furthermore,   as noted  in Ref.  \cite{Balitsky:2019krf},  the combination $G_{ij}G_{ij}$  (with summation over  transverse $i,j$) 
  has the same one-loop UV anomalous dimension
 as  ${M}_{0 i;  i 0 } $,  while the matrix element  $G_{30}G_{03}$ has the same one-loop UV anomalous dimension as ${M}_{3 i;  i 3 } $. 
 Hence,    the combinations of  Eqs. (\ref{00m}) and (\ref{33m}) are 
 multiplicatively renormalizable at the one-loop level.

\subsection{Reduced Ioffe-time distribution}

Within the pseudo-PDF approach \cite{Radyushkin:2017cyf}, the 
link-related UV divergences are eliminated through introducing
the reduced Ioffe-time distribution.  Namely, for each  multiplicatively  renormalizable amplitude $  \mathcal{M}$  we build the ratio 
 \begin{align}
{\mathfrak M} (\nu, z_3^2) \equiv \frac{{\mathcal{M}} (\nu, z_3^2)}{{\mathcal{M}}(0, z_3^2)} \  , 
 \label{redm0}
\end{align}
in which 
the   link-related   UV divergent $Z(z_3^2 \mu^2_{UV} )$  factors generated by the
vertex 
 and  link  self-energy diagrams 
cancel.  As a result, the small-$z_3^2$ dependence of the reduced 
pseudo-ITD  ${\mathfrak M} (\nu, z_3^2) $  comes from the logarithmic DGLAP evolution
effects only.

 \setcounter{equation}{0}  

\section{One loop corrections}

Below, we briefly summarize the results on ``non-box'' one-loop corrections
presented in Ref. \cite{Balitsky:2019krf}, and then discuss a rather lengthy  contribution of the box diagram 
that was presented there in part  only. 

\subsection{Link self-energy  contribution}

\begin{figure}[t]
   \centerline{\includegraphics[width=2in]{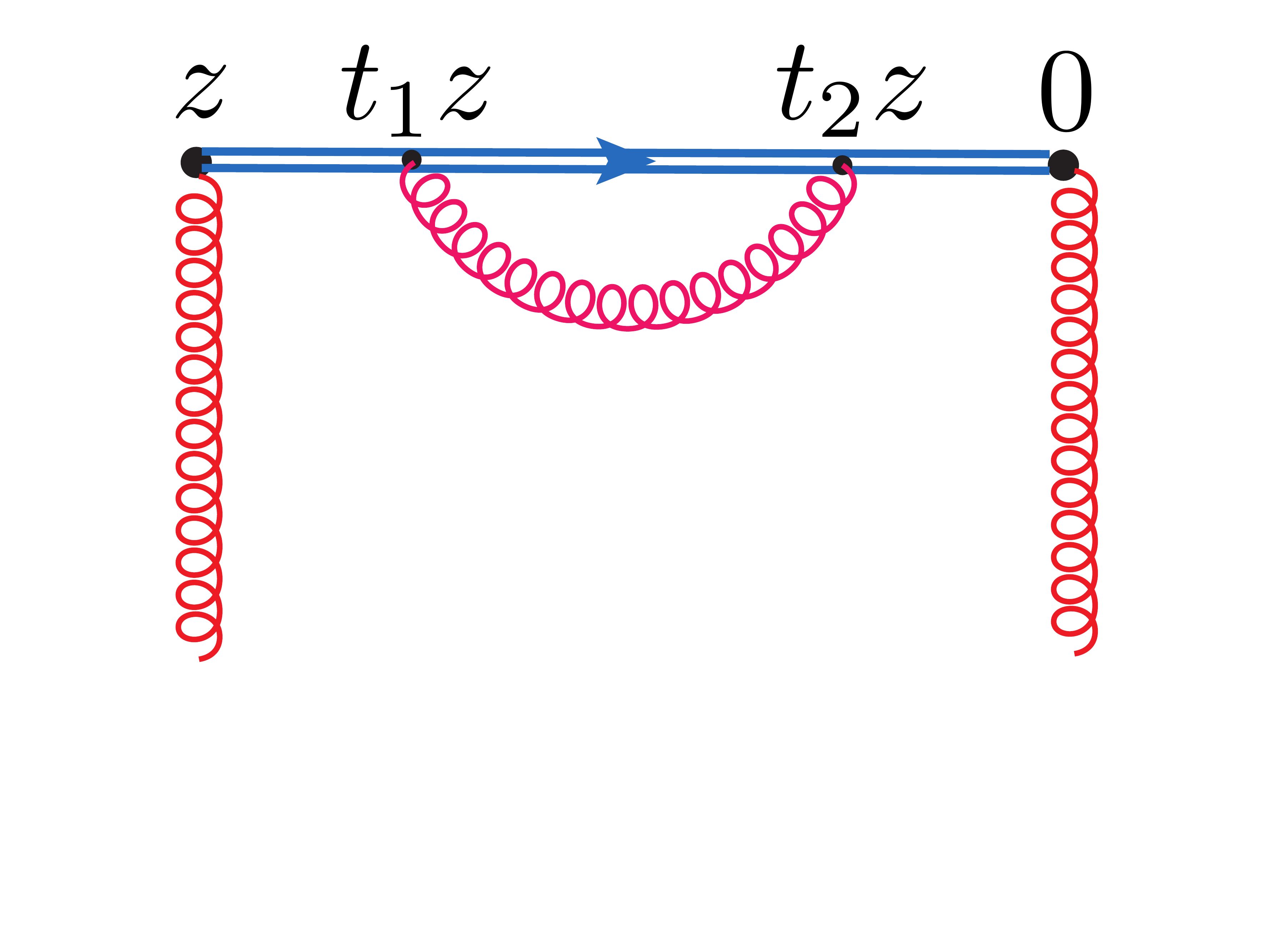}}
        \vspace{-8mm}
   \caption{Self-energy-type  correction for  the gauge link.
   \label{linkself}}
   \end{figure}

  The 
self-energy correction  for  the  gauge 
 link is given by the simplest diagram  (see Fig. \ref{linkself}). 
  In  lattice perturbation theory, it was 
  calculated 
at one loop in Ref. \cite{Chen:2016fxx}. 
An important property of this contribution is the presence of 
  a $\sim z_3/a_L$ linear term,
  where $a_L$ is  the lattice spacing 
  that provides here the   ultraviolet cut-off.

  Such corrections clearly factorize into a $\nu$-independent factor, and cancel in the ratio (\ref{redm0}),
  so that their explicit form is not essential in the pseudo-PDF approach. 
 Still, in dimensional regularization, one has 
 \begin{align}
&
-{g^2N_c \over 4\pi^2[(-z^2 \mu_{\rm UV}^2 +i\epsilon)]^{{d\over 2}-2}} {\Gamma\big({d/ 2}-1\big) \over (3-d)(4-d)}
\nn & \times G_{\mu\alpha}(z)G_{\lambda \beta }(0) \ ,
\label{selfAD}
\end{align}
where the pole for $d=3$ ($d=4$) corresponds to the linear (logarithmic)  UV divergences
present in this diagram.

\subsection{Vertex contribution} 

\label{vert} 

\begin{figure}[h]
   \centerline{\includegraphics[width=2.5in]{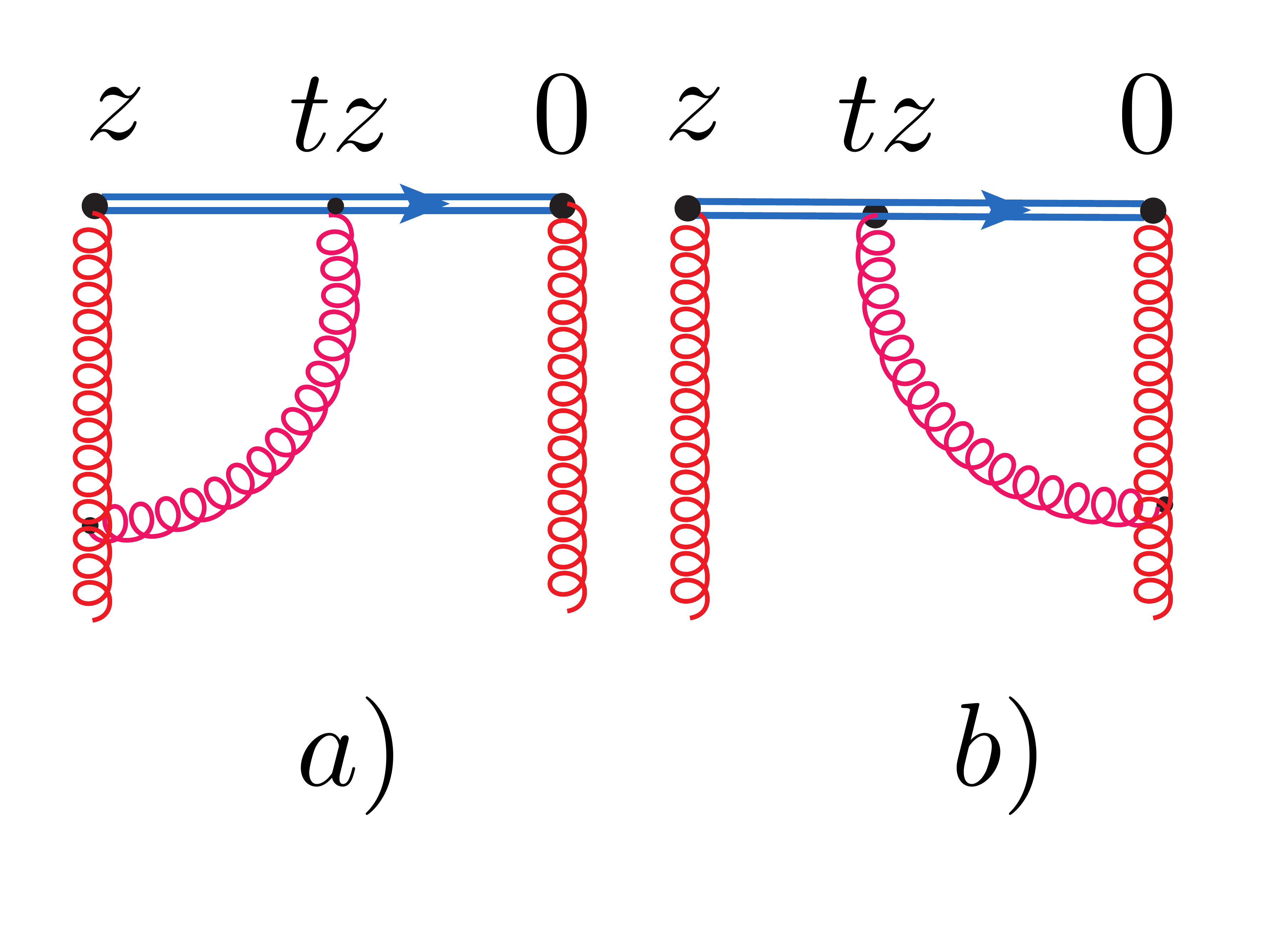}}
        \vspace{-5mm}
   \caption{Vertex diagrams with  gluons coming out of the gauge link.
   \label{link}}
   \end{figure}

There are   also vertex diagrams     
involving gluons  that connect the gauge link with the gluon  lines, see  \mbox{Fig. \ref{link}. }

We  use  the method of calculation 
described in Ref.  \cite{Balitsky:1987bk}. It is  based on the background-field technique, 
with the gluon propagator taken 
in the ``background-Feynman'' (bF) gauge \cite{Balitsky:1987bk}.  
The full, ``uncontracted''  vertex contribution  is given by 
\begin{align}
&  \mathcal{O}_{\mu\alpha;\nu\beta}^{\rm vertex } 
=  \, \frac{g^2N_c\Gamma(d/2-1)}{4\pi^2(-z^2)^{d/2-1}} \nn & \times \int_0^1 \dd u \left\{ \left(\frac{ u^{3-d}- u}{d-2} \right) G_{\mu\alpha}(\bar u z)\left(z_{\beta} G_{z\nu}(0) - z_{\nu} G_{z\beta}(0)\right) \right. \nn
&\quad \left. + \left(\frac{  u^{3-d}- u}{d-2} \right) \left(z_{\alpha} G_{z\mu}( \bar uz) - z_{\mu} G_{z\alpha}( \bar uz)\right) G_{\nu\beta}(0) \right\} \nn
&\quad   +\frac{g^2N_c\Gamma(d/2-2)}{8\pi^2(-z^2)^{d/2-2}} \nn & \quad \times  \int_0^1 \dd u \, 2\left[\frac{ u^{3-d}-1}{d-3}\right]_+   G_{\mu\alpha}( \bar uz)G_{\nu\beta}(0) \  . 
\label{vertexfull}
\end{align}

In this expression,   just  one of the fields in the 
$G_{\mu\alpha}(z)G_{\lambda \beta }(0)$ operator is corrected, while another remains intact. 
In particular, the diagram \ref{link}a changes $G_{\mu\alpha}(z)$ into the sum of two
terms.  One of them contains UV divergences, while the other one is UV finite. 
 The UV-divergent  term is   given by 
\begin{align}
&\frac{N_c g^2}{4\pi^2} 
\frac{\Gamma(d/2-1)}{(d-2 )(-z^2)^{d/2-1}}
\int_0^1 \dd u\,  \left(u^{3-d}-u\right)  \nn  &\quad  \times 
 \left(z_{\alpha} G_{z\mu}( \bar uz) - z_{\mu} G_{z\alpha}(\bar uz)\right) \ , 
 \label{2a1}
 \end{align}
 where $G_{z\sigma} \equiv z^\rho G_{\rho\sigma}$ and $\bar u \equiv 1-u$.  
 The overall
 \mbox{$d$-dependent}  factor  here is finite for $d=4$,  but  the \mbox{$u$-integral} diverges at the lower limit.
The divergence  disappears if one uses the UV    regularization by  taking 
 $d=4-2\varepsilon_{\rm UV}$,  which converts it 
 into a pole at $\euv=0$.

 Since the UV  divergence comes from the \mbox{$u \to 0 $}  integration, we 
 can isolate it by taking $\bar u =1$ in the gluonic field, which gives 
 \begin{align}
&\frac{N_c g^2}{8\pi^2} 
\frac{\Gamma(d/2-1)}{(-z^2)^{d/2-1}}
\frac1{4-d}  
 \left(z_{\alpha} G_{z\mu}(z) - z_{\mu} G_{z\alpha}(z)\right) \ . 
 \label{UVvert}
 \end{align}
 The remainder is given by 
 \begin{align}
&\frac{N_c g^2}{4\pi^2} 
\frac{\Gamma(d/2-1)}{(d-2 )(-z^2)^{d/2-1}}
\int_0^1 \dd u\,  \left[u^{3-d}-u\right]_{+(0)}  \nn  &\quad  \times 
 \left(z_{\alpha} G_{z\mu}( \bar uz) - z_{\mu} G_{z\alpha}(\bar uz)\right) \ , 
  \label{UVvertreg}
 \end{align}
 where the plus-prescription at $u=0$  is defined as
   \begin{align}
& \int_0^1  \dd u \left[f(u)\right]_{+(0)} g(u) = \int_0^1  \dd u f(u) [g(u) -g(0)]  \ .
 \label{plus0}
 \end{align}

The second, UV finite term from  the diagram \ref{link}a  is   given by 
 \begin{align}
\quad   \frac{N_c g^2}{8\pi^2} \frac{\Gamma(d/2-2)}{({d-3})(-z^2)^{d/2-2}} &
 \int_0^1 \dd u   \left[u^{3-d}-1\right]_{+(0)}  \nonumber 
 \\ &\quad  \times  
  G_{\mu\alpha}( \bar uz)  G_{\lambda \beta }( 0) \ .
\label{2aEv}
\end{align}
Note that the gluonic operator in Eq. (\ref{2aEv}) 
has the same tensor structure as the original operator 
$G_{\mu\alpha}( z)  G_{\beta \nu}( 0)$ differing from it just by 
rescaling $z \to \bar u z$. 
 There is no mixing with operators  of a different type. 
 The $u$-integral in this case does not diverge   for $d=4$, but the  overall 
\mbox{$\Gamma(d/2-2)$}  factor  has a pole  $1/(d-4)$.

Formally,  there is also a pole   $1/(d-3)$,   
corresponding to a linear UV divergence. 
However,  the singularity 
for $d=3$ is eliminated by the $\left [u^{3-d}-1\right ]$ combination in the integrand. 
One may say that the linear divergences present in ``$u^{3-d}$''  and ``$-1$''  
parts cancel each other.  

The remaining  $1/(d-4)$  pole  
corresponds to a collinear divergence 
developed 
 because  all  the propagators correspond to massless particles.

\subsection{Gluon self-energy diagrams}

\begin{figure}[t]
   \centerline{\includegraphics[width=1.5in]{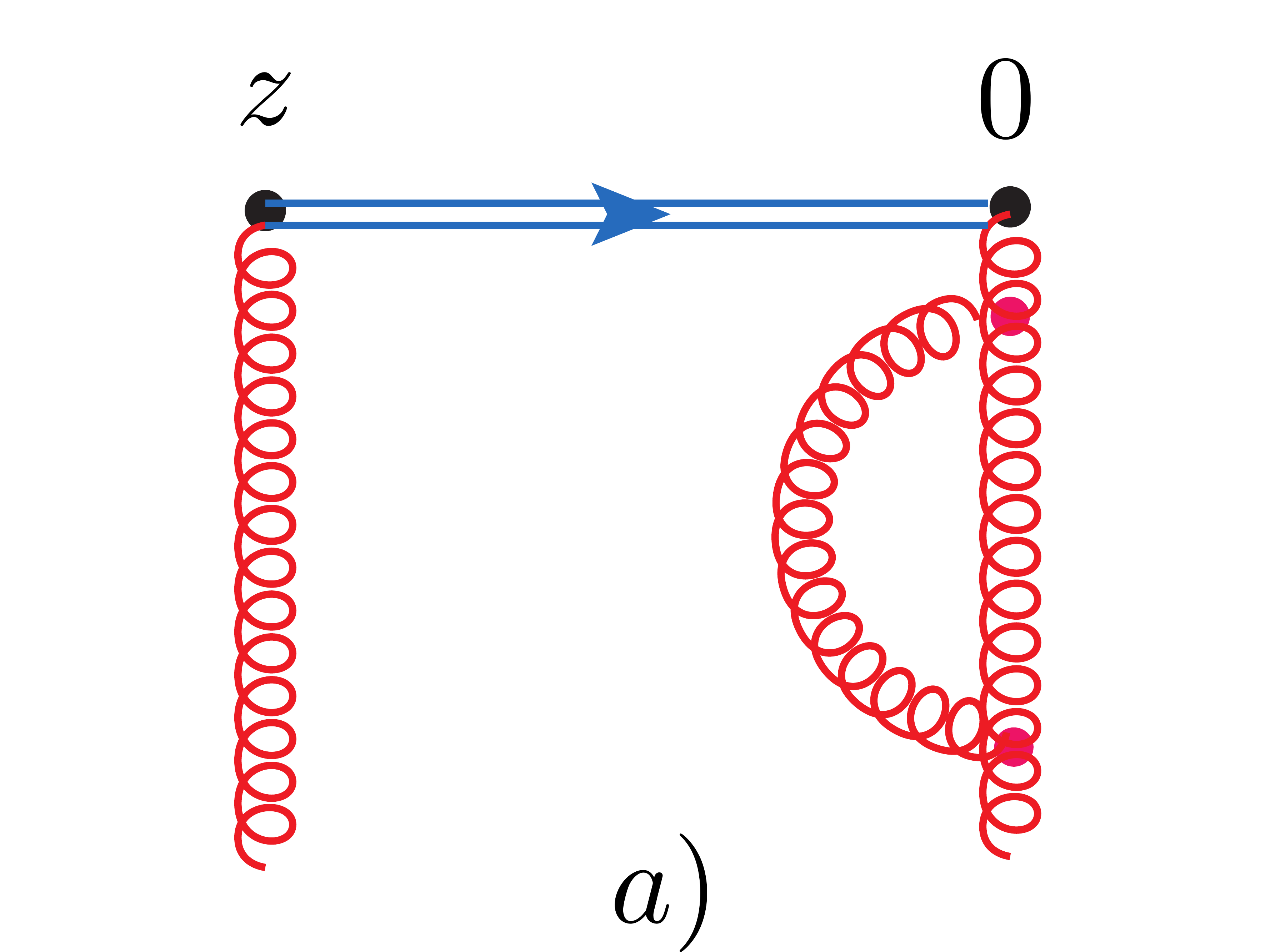} \hspace{-5mm} \includegraphics[width=1.5in]{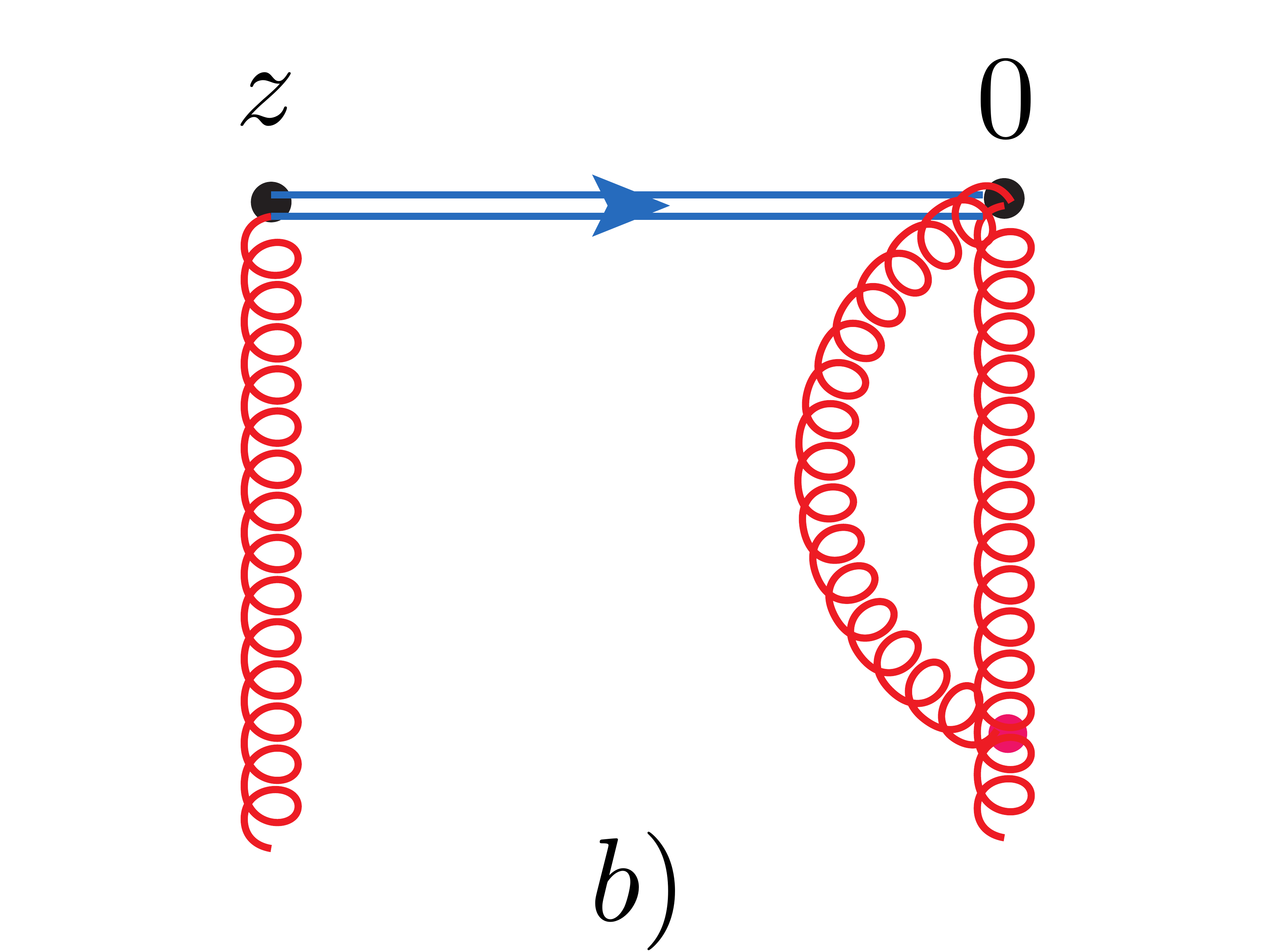}}
   \caption  { Gluon  self-energy-type insertions into the right leg.  
   \label{gluself}}
   \end{figure}

  Another simple type of one-loop corrections is represented by the gluon  self-energy diagrams,  one of which is shown in Fig. \ref{gluself}a.
These diagrams have both the UV and collinear 
divergences. 
The combined contribution 
of the Fig. \ref{gluself} diagrams and their left-leg analogs  is given by
\begin{align}
{g^2N_c\over 8\pi^2} 
\frac1{2-d/2}
\left [2 - \frac{\beta_0}{2N_c} \right ]G_{\mu\alpha}(z)G_{\lambda \beta }(0) \  ,
\label{counter3}
\end{align}
where $\beta_0 =11N_c/3$ in gluodynamics, so that the terms in the square bracket combine into 1/6.

\subsection{Box diagram}

  \begin{figure}[h]
   \centerline{\includegraphics[width=1.7in]{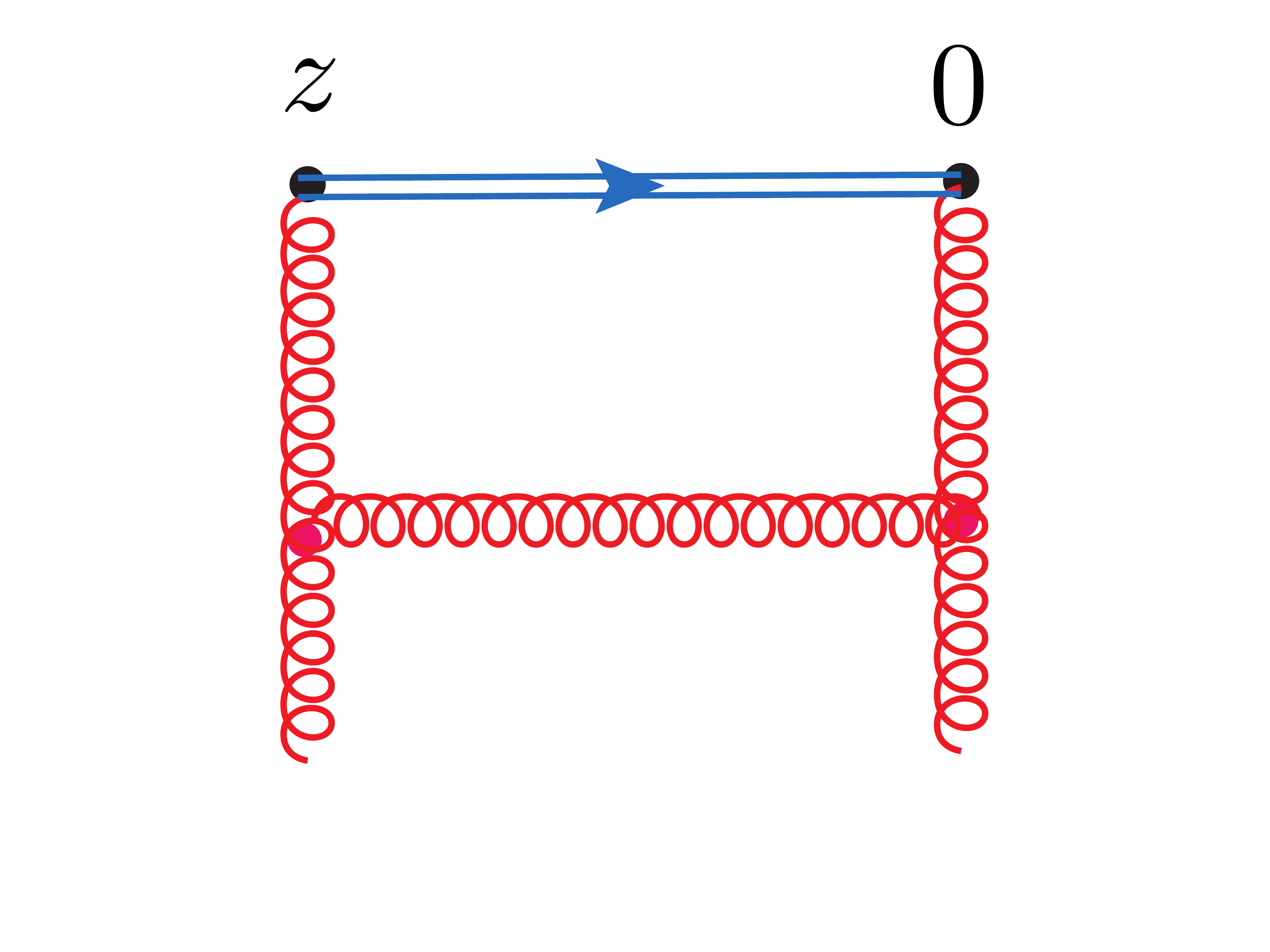}}
        \vspace{-5mm}
   \caption  {Box diagram.     \label{box}}
   \end{figure}

The most complicated technically is the calculation 
  of  the ``box'' diagram  which 
contains  a gluon exchange between
two gluon lines (see Fig.   \ref{box}).  
This diagram does not have UV divergences, but it has DGLAP 
$\log z_3^2$ contributions.  
In contrast to the vertex diagrams, the original  
$G_{\mu\alpha}( z)  G_{\nu \beta }( 0)$ operator generates 
now a mixture of bilocal operators corresponding 
to various projections of $G_{\mu\alpha}( uz)  G_{\nu \beta }( 0)$
onto the structures built from vectors $p$, $z$ and the metric tensor $g$. 

Our  result for arbitrary indices $\mu\alpha\nu \beta $ is given  in the Appendix A.
It is  presented  in the operator form,
however, it  contains only those operators that survive 
in the forward case, i.e., the operators that  have the form of a full derivative are abandoned. 
Still, the expression is rather lengthy. Furthermore, we mostly need it 
for particular combinations of indices corresponding to matrix elements 
 $ { M}_{0 i;  i 0 }  +  { M}_{j i ;  i j }$, $ { M}_{3 i;  i 3 }  +  2{ M}_{30 ;  03} $
and $ { M}_{0 i;   i 3 } +  { M}_{3 i;  i 0 } $ that  contain the $M_{pp}$ invariant amplitude and are   listed in Eqs. 
 (\ref{00m}), (\ref{33m}) and (\ref{03m}).
To shorten the formulas, let us introduce the following notations for the  bilocal operators corresponding to these matrix elements.  
  \begin{widetext}
\begin{align} 
{\mathcal O}_{00 } (z_1,z_2) = &G_{ 0  i }(z_1)  G_{  i 0 }(z_2) + G_{  i j  }(z_1)G_{  j i }(z_2)  \  , \\
{\mathcal O}_{33 } (z_1,z_2) = &G_{ 3  i }(z_1)  G_{  i 3 }(z_2) + 2G_{  30  }(z_1)G_{  03 }(z_2)  \  ,  \\
{\mathcal O}_{03 }^+ (z_1,z_2) = &G_{ 0  i }(z_1)  G_{  i 3 }(z_2) + G_{  3i }(z_1)G_{ i0}(z_2)  \  .
\label{3comb}
\end{align}

In the case of  ${\mathcal O}_{00 } (z,0) $ and ${\mathcal O}_{33 }(z,0) $ operators, the box diagram produces the following corrections 
\begin{align}
\label{00p}
{\mathcal O}_{00 } (z,0)   \stackrel{\rm box}{\longrightarrow}  & \, - \frac{g^2N_c\Gamma(d/2-1)}{4\pi^2\left(-z^2\right)^{d/2-2}}   \int_0^1 \dd u \,   \left ( u\bar u +\frac {\bar u^3}{3}\right )
 {\mathcal O}_{33 } (uz,0)   \nn
&\quad + \frac{N_c\Gamma(d/2-2)}{8\pi^2\left(-z^2\right)^{d/2-2}} \int_0^1 \dd u   \left\{\vphantom{\frac{1}{2}} \right.      
   \left( \bar u \left( u^2+1 \right) - 2 u\right)   {\mathcal O}_{00 } (uz,0) 
+     \bar u \left( u^2+1 \right)
  {\mathcal O}_{33 } (uz,0)    \left. \vphantom{\frac{1}{2}} 
  \right\}      \ , 
\end{align}
\begin{align}
\label{33p}
{\mathcal O}_{33 } (z,0)   \stackrel{\rm box}{\longrightarrow} & \, - \frac{g^2N_c\Gamma(d/2-1)}{4\pi^2\left(-z^2\right)^{d/2-2}}   \int_0^1 \dd u  \left\{ \vphantom{\frac{1}{2}}    \left (\bar u +u\bar u +\frac{\bar u^3}{3}\right )
 {\mathcal O}_{33 } (uz,0) 
 -  \bar u    {\mathcal O}_{00 } (uz,0) 
  \right\}\nonumber \\
& \quad + \frac{N_c\Gamma(d/2-2)}{8\pi^2\left(-z^2\right)^{d/2-2}} \int_0^1 \dd u   \left\{\vphantom{\frac{1}{2}} \right.   
  \left( \bar u \left( u^2+ 1 \right) -2u \right)   {\mathcal O}_{33 } (uz,0)   
+ \bar u \left( u^2+ 1 \right)   {\mathcal O}_{00 } (uz,0)      \left. \vphantom{\frac{1}{2}} \right\}  \  .
\end{align}
%


 One can see  that 
the box diagram contribution for each of them involves 
matrix elements of both  the operators ${\mathcal O}_{00 } (uz,0) $ and ${\mathcal O}_{33 } (uz,0) $.
Thus,  
 these  two operators  mix here with each other.   
Furthermore, matrix elements of both of them contain the $ \mathcal{M}_{pp}$ invariant amplitude.
Thus, it is interesting to rewrite Eqs. (\ref{00p}) ,  (\ref{33p}) 
in terms of the invariant functions:
\begin{align}
\label{00pm}
 \mathcal{M}_{pp} (\nu , z^2 ) &    \stackrel{\rm box}{\longrightarrow}   \, - \frac{g^2N_c\Gamma(d/2-1)}{4\pi^2\left(-z^2\right)^{d/2-2}}  
  \int_0^1 \dd u  \,  \left (u\bar u+\frac{\bar u^3}{3} \right )
 \left[  \mathcal{M}_{pp} (u\nu , z^2 )- z_3^2 \mathcal{M}_{ppzz} (u\nu , z^2 )\right]  \nn
&\quad + \frac{g^2N_c\Gamma(d/2-2)}{8\pi^2\left(-z^2\right)^{d/2-2}} \int_0^1 \dd u   \left\{\vphantom{\frac{1}{2}} \right.        2 \left( \bar u \left( u^2+1 \right) - u\right)  \,   \mathcal{M}_{pp} (u\nu , z^2 ) 
-  \bar u \left( u^2+1 \right)  z_3^2 \mathcal{M}_{ppzz}  (u\nu , z^2 )     \left. \vphantom{\frac{1}{2}} \right\}   \ , 
\end{align}
\begin{align}
\label{33pm}
 \mathcal{M}_{pp} (\nu , z^2 )- z_3^2 \mathcal{M}_{ppzz} (\nu , z^2 )   \stackrel{\rm box}{\longrightarrow}  & \, - \frac{g^2N_c\Gamma(d/2-1)}{4\pi^2\left(-z^2\right)^{d/2-2}}   \int_0^1 \dd u  \left\{ \vphantom{\frac{1}{2}}    \left (u\bar u+\frac{\bar u^3}{3}\right  )     \mathcal{M}_{pp} (u\nu , z^2 )
  \right. \nn &\quad \quad  \quad  \quad  \quad  \quad \quad \quad \quad \quad  \quad  \quad  \left.  - 
  \left (u\bar u+\frac{\bar u^3}{3}+\bar u \right) z_3^2 \mathcal{M}_{ppzz} (u\nu , z^2 ) 
   \right\}\nn
& \quad + \frac{g^2N_c\Gamma(d/2-2)}{8\pi^2\left(-z^2\right)^{d/2-2}} \int_0^1 \dd u   \left\{\vphantom{\frac{1}{2}} \right.  
  2 \left( \bar u \left( u^2+ 1 \right) -u \right)      \mathcal{M}_{pp} (u\nu , z^2 )  
    \nn &\quad \quad  \quad  \quad  \quad  \quad \quad  \quad  \quad   \quad  \quad  \quad \quad 
     -[ \bar u \left( u^2+ 1 \right)   -2u ] \, z_3^2 \mathcal{M}_{ppzz} (u\nu , z^2 )   \left. \vphantom{\frac{1}{2}}
      \right\} \ .
\end{align}
These relations have a very similar structure and, in fact,  coincide if one discards the $ \mathcal{M}_{ppzz}$  terms. 
Taking the difference of these expressions gives a very simple result for the box correction to $ \mathcal{M}_{ppzz}$
\begin{align}
\label{ppzz}
 \mathcal{M}_{ppzz} (\nu , z^2 )   \stackrel{\rm box}{\longrightarrow}  & \, - \frac{g^2N_c\Gamma(d/2-1)}{4\pi^2\left(-z^2\right)^{d/2-2}}   \int_0^1 \dd u  
\, 
\bar u \,  \mathcal{M}_{ppzz} (u\nu , z^2 ) 
    - \frac{g^2N_c\Gamma(d/2-2)}{4\pi^2\left(-z^2\right)^{d/2-2}} \int_0^1 \dd u   
   \, u \,  \mathcal{M}_{ppzz} (u\nu , z^2 ) 
     \ .
\end{align}

The situation is simpler for the ${\mathcal O}_{03 }^+(z,0)$ operator,  for   
which   the box diagram contribution
 is expressed through the ${\mathcal O}_{03 }^+(uz,0)$ operator 
 only, 
\begin{align}
\label{03p}
{\mathcal O}_{03 }^+(z,0)   \stackrel{\rm box}{\longrightarrow}  & \,  \frac{g^2N_c\Gamma(d/2-1)}{4\pi^2\left(-z^2\right)^{d/2-2}}   \int_0^1 \dd u \,  \left( u  \bar u + \frac23{\bar u^3} \right)    {\mathcal O}_{03 }^+(uz,0)     \nn
&+   \frac{N_c\Gamma(d/2-2)}{4\pi^2\left(-z^2\right)^{d/2-2}} \int_0^1 \dd u   \,    \left( \bar u \left(u^2+1 \right)  - u\right)    {\mathcal O}_{03 }^+(uz,0)   \ .
\end{align}

In all the  cases, Eqs. (\ref{00pm}), (\ref{33pm}), and (\ref{03p}), the $\Gamma(d/2-2)$ terms are singular for $d=4$, which results  in 
$\log (-z^2)$  terms generating the DGLAP evolution.
The $\Gamma(d/2-1)$  terms are singular for $d=2$, which corresponds 
to the fact that the gluon propagator in two  dimensions 
has a logarithmic $\log (-z^2)$ behavior in the coordinate space.
For $d=4$, these terms are finite.
Note that, unlike the vertex part, the box  contribution 
does not have the plus-prescription form.

\setcounter{equation}{0}

\section{DGLAP evolution structure}

Adding the results for all the diagrams discussed above, we get the following expressions 
for their combined contribution for the 3 operator combinations listed  in Eq. (\ref{3comb})
\begin{align}
\label{30full}
&   { M}_{0 i; i 3 } + { M}_{3 i;i 0 } = 4 p_0     p_3  \mathcal{M}_{pp} \left(\nu,z_3^2\right)+ 2 p_0 z_3  \left ( \mathcal{M}_{pz} \left(\nu,z_3^2\right)+\mathcal{M}_{zp} \left(\nu,z_3^2\right)\right) \nn
& \to \frac{g^2N_c}{8\pi^2}\int_0^1 \dd u  
 \left\{ \left[ \left (\frac{3}{2} -\frac16 \right )  \log(z_3^2 \mu_{\rm UV}^2 \frac{e^{2\gamma_E}}{4})  +2 \right] \delta(\bar u)  +    \left [ u -3  \frac{u}{ \bar u} -4 \frac{ \log (\bar u)}{\bar u}     \right ]_+ +         2 \left  (\bar uu +  \frac{2}{3}\bar u^3  \right ) \right. \nn
 &\quad  \left.       -        2  \log(z_3^2  \mu_{\rm IR}^2 \frac{e^{2\gamma_E}}{4}) \left [ \frac{ (1- u \bar u)^2}{\bar u} \right ]_+  \right\}  \, \left( 4 p_0     p_3  \mathcal{M}_{pp}\left(u\nu,z_3^2 \right) + 2 u p_0 z_3  \left ( \mathcal{M}_{pz} \left(u\nu,z_3^2 \right)+\mathcal{M}_{zp} \left(u\nu,z_3^2 \right)\right) \right)   \ , 
\end{align}
\begin{align}
\label{00full}
& \frac{M_{0i;i0}+ M_{ i j ; j i} }{2p_0^2} =   \mathcal{M}_{pp} \left(\nu,z_3^2\right) \nn
&\to   \frac{g^2N_c}{8\pi^2} \int_0^1 \dd u  \left\{ \left( \left(1 -{1 \over 6}\right)\log\left(z_3^2\mu^2_{\rm UV } \frac{e^{2\gamma_E}}{4}\right) +2 \right) \delta\left( \bar u \right)    -    \left({1 \over 2} \delta\left(\bar u \right) +  \left[{2 \over 3 } \left(1-u^3\right)+ {4 u+4 \log (\bar u) \over \bar u}\right]_+\right)  \right. \nn
  &\quad \left. -  \log \left(z_3^2\mu^2_{\rm IR}\frac{e^{2\gamma_E}}{4}\right)    \left[ {2 \left(1- u \bar u\right)^2 \over \bar u}\right]_+ \right\} \,   \mathcal{M}_{pp} \left(u\nu,z_3^2 \right)    \nn
    &\quad + \frac{g^2N_c}{8\pi^2}   \int_0^1 \dd u   \left\{  {2 \over 3 } \left(1-u^3\right)  + \log \left(z_3^2\mu^2_{\rm IR}\frac{e^{2\gamma_E}}{4}\right)     \bar u \left( u^2+1 \right) \right\} \,  u^2 z_3^2  \mathcal{M}_{ppzz} \left(u\nu,z_3^2 \right)  \ , 
    \end{align}
\begin{align}
\label{33full}
& \frac{M_{3i;i3}+ 2 M_{ 3 0 ; 0 3 } }{  2  p_0^2} =  \mathcal{M}_{pp} \left(\nu,z_3^2\right) -  z_3^2 \mathcal{M}_{ppzz} \left(\nu,z_3^2\right) \nn
& \to \frac{g^2N_c}{8\pi^2} \int_0^1 \dd u \left\{ \left( \left(2 - {1 \over 6} \right) \log\left(z_3^2 \mu^2_{\rm UV}\frac{e^{2\gamma_E}}{4}\right) +2 \right)\delta\left( \bar u \right)  -   \left(  {1 \over 2 }  \delta \left( \bar u \right)+ \left[ {2 \over 3 } \left( 1-u^3 \right) +  {2u^2 + 4 \log (\bar u) \over \bar u}\right]_+ \right) \right.   \nn
&\quad  \left. - \log \left(z_3^2\mu^2_{\rm IR}\frac{e^{2\gamma_E}}{4}\right)  \left[ {2 \left(1-u\bar u\right)^2 \over \bar u} \right]_+ \right\} \left(  \mathcal{M}_{pp} \left(u\nu,z_3^2 \right) -  u^2 z_3^2 \mathcal{M}_{ppzz} \left(u\nu,z_3^2 \right) \right) \nn
&\quad -\frac{g^2N_c}{8\pi^2}   \int_0^1 \dd u \left\{   2 \bar u     +\log (z_3^2 \mu_{\rm IR}^2\frac{e^{2\gamma_E}}{4})    \bar u \left( u^2+ 1 \right)  \right\}  u^2 z_3^2 \mathcal{M}_{ppzz} \left(u\nu,z_3^2 \right)   \ .
\end{align}
All these  combinations contain  the $\log \left(z_3^2\mu^2_{\rm IR}\frac{e^{2\gamma_E}}{4}\right) $ evolution term
accompanied by the  $gg$-component of the  Altarelli-Parisi (AP) kernel 
  \begin{align} 
B_{gg} (u)  = 
     2  \left [\frac{(1-u\bar u)^2} {\bar u }   \right ]_{+} \ .
     \label{V1}
  \end{align}   
However, they have  different $z_3$-independent parts,    as a result of mixing with ``higher-twist''  functions
  $\mathcal{M}_{zp}+ \mathcal{M}_{pz}$ in Eq. (\ref{30full}), and $\mathcal{M}_{ppzz}$ in Eqs. (\ref{00full}) and (\ref{33full}).
   The kernel  (\ref{V1})  has  the plus-prescription 
 structure reflecting the fact that, 
in the local limit,  ${\mathcal{M}}_{pp} (z,p)$  is proportional to   the  
matrix element of the gluon energy-momentum tensor that is conserved in the absence of the gluon-quark interactions.
From now on,  ``+'' means the plus-prescription at 1.

The $\log \left(z_3^2\mu^2_{\rm UV}\frac{e^{2\gamma_E}}{4}\right) $ term in each result comes from the UV-singular 
contributions. They contain the $\delta(\bar u) $ factor
which reflects the local nature of the UV divergences and converts $M(uz,p)$ into $M(z,p)$. Each result shares the same UV-singular contribution from the link renormalization and self energy contributions, but differ in their vertex contribution, as mentioned in section 
\ref{vert}.

The expressions given above include gluon-gluon transitions only.
Thus, we need to include also the one-loop diagrams describing the gluon-quark transition.

\subsection{Gluon-quark mixing}

  \begin{figure}[h]
   \centerline{\includegraphics[width=1.7in]{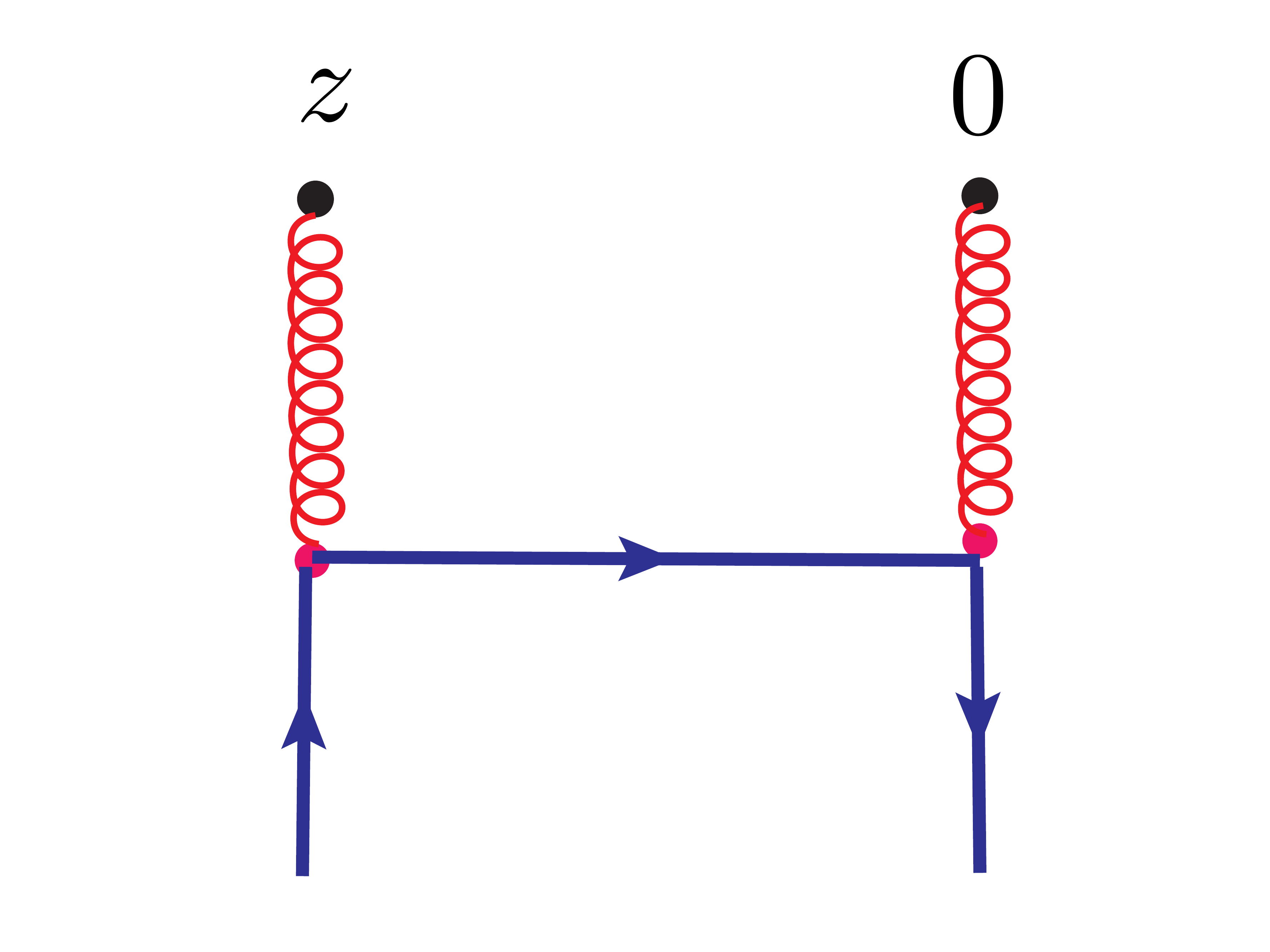}}
        \vspace{-3mm}
   \caption  {Gluon-quark mixing diagram.     \label{gluqum}}
   \end{figure}

The correction to   the  gluon operator with arbitrary indices generated by  the gluon-quark diagram shown in Fig. \ref{gluqum} 
is presented in Appendix B. To illustrate its structure, let us take the projection corresponding to the ${\mathcal O}_{03 }^+ $
operator.  In the $\overline{\rm MS}$ scheme,  it reads
 \begin{align}
\frac{ g^2 C_F  }{4\pi^2 z_3}&   \int_0^1 \dd u\left[  2\bar u \mathcal{O}^0_q \left( uz_3 \right)  + \left. 2\bar u u  z_3 \partial_0    \mathcal{O}^3_q \left( uz \right) \right|_{z_0 = 0}  \right] \nn
&-   \frac{ g^2 C_F}{4\pi^2}  \ln \left( z_3^2 \mu^2_{IR} \frac{e^{2\gamma_E}}{4}  \right)
\int_0^1 \dd u 
  \left[   (2\bar uu +\bar u^2)\partial_3 \mathcal{O}^0_q \left( uz_3 \right)   -(\bar u^2 +u^2) \left. \partial_0 \mathcal{O}^3_q \left( uz \right)\right|_{z_0 = 0}  \right]   \  . 
  \label{gqmix}
\end{align}  
  The  
 singlet combination of quark fields  $ {\cal O}_q^\mu (z_3)$  is defined as  
  \begin{align}
& {\cal O}_q^\mu (z)=  \frac{i}{2} \sum_f  \left(\bar\psi_f( z)  \gamma^\mu  \psi_f (0) - \bar\psi_f( 0)  \gamma^\mu  \psi_f(z)  \right)  \ ,
 \label{gsing}
\end{align}  
with $f$ numerating  quark  flavors. 
Since the matrix element of ${\cal O}_q^\mu ( z)$ is odd in $z$, it can be parametrized by 
 \begin{align}
 \langle p| {\cal O}_q^\mu  ( z_3) |p\rangle
= & 2p^\mu \int_0^1 \dd y \sin\left( y pz \right) f_S (y ) 
=  -2  p^\mu  \nu
\int_0^1 \dd \alpha\, {\cal I}_S (\alpha \nu) \ , 
 \label{qpar2}
\end{align}  
where $\nu = -(pz) $,  as usual, and   
 \begin{align}
&  {\cal I}_S (\nu)  =
\int_0^1 \dd y \ \cos (y   \nu) \,  y\,    f_S (y) 
 \label{MS}
\end{align}  
is the singlet quark Ioffe-time distribution. 

Applying this parametrization, the $gq$ correction to $M_{03} ^+$ may be written as
\begin{align}
&\bra{p} G_{3i}(z) G_{i0}(0)+G_{0i}(z) G_{i3}(0) \ket{p} 
\quad \to - 2p_0 p_3 \frac{ g^2 C_F }{4\pi^2}   \int_0^1 \dd u \,
  \left [    \ln \left( z_3^2 \mu^2_{IR}\frac{e^{2\gamma_E}}{4} \right)   {\cal B}_{gq} (u) + \bar u (1+u)    \right ] \mathcal{I}_S \left(u\nu \right)  \ , 
\end{align}
with the $gq$ component of the evolution kernel given by $ {\cal B}_{gq} (u) \equiv 1+(1-u)^2$. For  two  other matrix elements
listed in Eq. (\ref{3comb}), the  analogs of Eq. (\ref{gqmix}) are given by
\begin{align}
& \bra{p}G_{0i}\left( z \right) G_{i0} \left( 0 \right) \ket{p} + \bra{p}G_{ij}\left( z \right) G_{ji} \left( 0 \right) \ket{p}\to  -    p_0^2 
 \frac{g^2 C_F   }{4\pi^2}    \ln \left( z_3^2 \mu^2_{IR}\frac{e^{2\gamma_E}}{4} \right) 
\int_0^1 \dd u \,  {\cal B}_{gq} (u)  \mathcal{I}_S \left( u\nu \right)  
\label{00gq}
\end{align}
and 
\begin{align}
&\bra{p}G_{ 3  i }\left( z \right) G_{   i3 } \left( 0 \right) \ket{p} + 2\bra{p} G_{30}\left( x \right) G_{03} \left( 0 \right) \ket{p}
 \quad \to 
-   p_0^2   \frac{ g^2 C_F }{4\pi^2}   \int_0^1 \dd u \,  \left [    \ln \left( z_3^2 \mu^2_{IR}\frac{e^{2\gamma_E}}{4} \right)   {\cal B}_{gq} (u) + 4 \bar u   \right ] \mathcal{I}_S \left(u\nu \right)  \ .
\label{33gq} 
\end{align}


\subsection{Matching relations}

As discussed already, 
the combination 
$
 { M}_{0 i;  i 0 }  +  { M}_{ i j; j i} =   2   p_0^2  \mathcal{M}_{pp}   $,    at  the tree level,  is  proportional 
to the twist-2 amplitude $ \mathcal{M}_{pp}$ without contaminations. 
 The amplitude  $ \mathcal{M}_{pp} (\nu, z_3^2)$ obtained in this way
 may be used 
  to form the reduced pseudo-ITD
   \begin{align}
 {\mathfrak  M}_{00}(  \nu,z_3^2) \equiv \frac{{\mathcal{M}_{pp}} (\nu, z_3^2)}{{\mathcal{M}_{pp}}(0, z_3^2)} \  
 \label{redm00}
\end{align}
 as in Eq. (\ref{redm0}).  
 Using the   results (\ref{00full}),  (\ref{00gq}) of our  calculations for the one-loop corrections to
 $
 { M}_{0 i; i 0 }  +  { M}_{ i j; j i}$, 
  we obtain 
 the matching  relation 
\begin{align}
 {\mathfrak M}_{00} (\nu, z_3^2)\,   {{\cal I}_g (0, \mu^2) } =&   { {\cal I}_g (\nu, \mu^2) }
\nn &-   \frac{\alpha_s N_c}{2\pi}\int_0^1 \dd u \,  {{\cal I}_g (u \nu, \mu^2)}
 \Biggl \{   \log (z_3^2  \mu^2 \frac{e^{2\gamma_E}}{4})  \,
 B_{gg} (u) 
          +    4 \left [  \frac{ u + \log (\bar u)}{\bar u}\right ]_{+}    +  \frac23   \left [ 1 -u^3
     \right ]_+ 
   \Biggr \}  
        \nonumber \\
 &
   -   \frac{\alpha_s C_F}{2\pi}   \log (z_3^2  \mu^2 \frac{e^{2\gamma_E}}{4}) 
 %
   \int_0^1 \dd u \, 
   \left  [  {\cal I}_S (u \nu, \mu^2)-  {\cal I}_S (0, \mu^2) \frac{{\cal I}_g (\nu, \mu^2)}{{\cal I}_g (0, \mu^2)} 
   \right ] 
 \, {\cal B}_{gq} (u) \ 
   \nn
    &\quad + \frac{\alpha_s N_c}{2\pi}   \int_0^1 \dd u   \left\{  {2 \over 3 } \left(1-u^3\right)  + \log \left(z_3^2\mu^2_{\rm IR}\frac{e^{2\gamma_E}}{4}\right)     \bar u \left( u^2+1 \right) \right\} \, 
       \nonumber \\
 & \quad  \times 
     u^2 z_3^2 \left [ \mathcal{M}_{ppzz} \left(u\nu,z_3^2 \right) - \mathcal{M}_{ppzz} \left(0,z_3^2 \right)  \frac{{\cal I}_g (\nu, \mu^2)}{{\cal I}_g (0, \mu^2)} \right ]
  \label{matching} 
\end{align}
between   the ``lattice function''   $ {\mathfrak M}_{00} (\nu, z_3^2)$
and the light-cone ITDs  $  {\cal I}_g (\nu,\mu^2)$  and 
 $  {\cal I}_S(\nu,\mu^2)$.  
  
 This matching condition also includes the  ``higher  twist'' term ${\cal M}_{ppzz}$ on its right-hand side.
 This term 
is accompanied by a $z_3^2$ factor that suppresses  its contribution  for small $z_3^2$ values,
and is omitted  in the matching conditions given in our original paper  \cite{Balitsky:2019krf}.
The size of the ${\cal M}_{ppzz}$ term may be estimated by comparing the lattice signals for 
 $
 { M}_{0 i;  i 0 }  +  { M}_{ i j; j i} $ and $M_{3i;i3}+2 M_{ 3 0 ; 0 3 }$   matrix elements.
 To this end, denoting  $ {M_{3i;i3}(z_3,p) + 2M_{ 3 0 ; 0 3 }(z_3,p) }\equiv {2p_0^2}{\mathcal{M}_{33}} (\nu, z_3^2)$,   
 we define  the ``33'' reduced ITD,
  \begin{align}
 {\mathfrak  M}_{33}(  \nu,z_3^2) \equiv \frac{{\mathcal{M}_{33}} (\nu, z_3^2)}{{\mathcal{M}_{33}}(0, z_3^2)} \  .
 \label{redm00}
\end{align}
 Now, using Eqs. (\ref{33full}) and  (\ref{33gq})
 we obtain  the  matching condition 
 \begin{align}
 {\mathfrak M}_{33} & (\nu, z_3^2)\, \Big [ {{\cal I}_g (0, \mu^2) }  -  z_3^2 \mathcal{M}_{ppzz} \left(0,z_3^2 \right)
 \Big ] = { {\cal I}_g (\nu, \mu^2) }
 -  z_3^2 \mathcal{M}_{ppzz} \left(\nu,z_3^2\right) \nn  & 
\hspace{-2mm}-   \frac{\alpha_s N_c}{2\pi}\int_0^1 \dd u \,  \Big [{{\cal I}_g (u \nu, \mu^2)} - 
u^2  z_3^2 \mathcal{M}_{ppzz} \left(u\nu,z_3^2 \right)\Big ]
 \Biggl \{     \log (z_3^2  \mu^2 \frac{e^{2\gamma_E}}{4})  \, B_{gg} (u) +4 \left [  \frac{ u^2/2 + \log (\bar u)}{\bar u}\right ]_{+}    +  \frac23   \left [ 1 -u^3
     \right ]_+  \Biggr \}  
      \nonumber \\
 &
   -   \frac{\alpha_s C_F}{2\pi}  
 %
   \int_0^1 \dd u \, 
    \left [{\cal I}_S (u\nu, \mu^2) - {\cal I}_S (0, \mu^2) \frac{{{\cal I}_g (\nu, \mu^2) }  -  z_3^2 \mathcal{M}_{ppzz} \left(\nu,z_3^2 \right)}{{{\cal I}_g (0, \mu^2) }  -  z_3^2 \mathcal{M}_{ppzz} \left(0,z_3^2 \right)}\right ]
 \,   \left \{ \log (z_3^2  \mu^2 \frac{e^{2\gamma_E}}{4}) {\cal B}_{gq} (u) +4 \bar u  \right \} \ 
   \nn
    &\hspace{3cm}- \frac{\alpha_s N_c}{2\pi}   \int_0^1 \dd u   \left\{ 2 \bar u   + \log \left(z_3^2\mu^2_{\rm IR}\frac{e^{2\gamma_E}}{4}\right)     \bar u \left( u^2+1 \right) \right\} \, \nn
    &\hspace{3cm}  \times  u^2 z_3^2 \left [ \mathcal{M}_{ppzz} \left(u\nu,z_3^2 \right) - \mathcal{M}_{ppzz} \left(0,z_3^2 \right)  \frac{{{\cal I}_g (\nu, \mu^2) }  -  z_3^2 \mathcal{M}_{ppzz} \left(\nu,z_3^2 \right)}{{{\cal I}_g (0, \mu^2) }  -  z_3^2 \mathcal{M}_{ppzz} \left(0,z_3^2 \right)}\right ]
  \label{33matching} 
\end{align}
that may be combined with  Eq. (\ref{matching}) to estimate the impact of the $ \mathcal{M}_{ppzz}$
contamination.

 The gluon  light-cone ITD  $  {\cal I}_g (\nu,\mu^2)$    is 
 related to the  gluon PDF  ${f}_g(x,\mu^2)$
by 
   \begin{align}
  {\cal I}_g (\nu,\mu^2) =\frac12 
   \int_{-1}^1 dx \,  \, 
e^{ix\nu} \,x  {f}_g(x,\mu^2)  \   .  
 \label{If}
\end{align}
In fact, $x  {f}_g(x,\mu^2) $ is an even function of $x$. Hence, 
the real part of $  {\cal I}_g (\nu,\mu^2)$ is given by the cosine transform of $x  {f}_g(x,\mu^2) $,
while its imaginary part vanishes. 
The overall factor ${\cal I}_g (0, \mu^2) $ corresponds to  the  fraction of the hadron 
 momentum carried by the gluons, ${\cal I}_g (0, \mu^2)  =  \langle x_g \rangle_{\mu^2}$. \
 This means that  Eq. (\ref{matching})  allows to extract just the shape of  the gluon distribution.
 Its normalization, i.e., the magnitude  of $ \langle x_g \rangle_{\mu^2} $  must  be 
 taken from  an independent  lattice calculation, similar to that performed 
 in Ref. \cite{Yang:2018bft}.
 The singlet quark function ${\cal I}_S (w \nu, \mu^2)$ that appears in the  
 ${\cal O} (\alpha_s)$ correction should be also calculated (or estimated)     
 independently.  
 
The matching condition (\ref{matching}) (without the $\mathcal{M}_{ppzz}$ terms and neglecting  
the gluon-quark mixing term) has been already used 
in lattice extractions of the unpolarized gluon PDFs 
by the MSU group \cite{Fan:2020cpa,Fan:2021bcr}  and the HadStruc collaboration \cite{HadStruc:2021wmh}.

\section{ Summary.}

In  this paper, we have presented the results that form the basis for 
the ongoing  efforts to calculate gluon PDF using the pseudo-PDF approach.

In particular, we have displayed our results  for the most complicated box diagram.
We have presented the expression for  the situation when all four indices are arbitrary,
and also for combinations of indices corresponding to three matrix elements
that are most convenient to extract the twist-2 invariant amplitude ${\cal M}_{pp}$. 
We also displayed the evolution structure for these matrix elements.

The results of our earlier publication  \cite{Balitsky:2019krf,Balitsky:2021bds} have been already 
used in the lattice extractions \cite{Fan:2020cpa,Fan:2021bcr,HadStruc:2021wmh}  of the gluon PDF
from the studies of the $ { M}_{0 i;  i 0 }  +  { M}_{j i ;  i j }$ matrix element.
The additional results for the box diagram and the gluon-quark  contribution given in the present
paper may be used for extractions of the gluon PDF from two other matrix elements,
with a possible cross-check of the results obtained from different matrix elements.

{\bf Acknowledgements.} We  thank K. Orginos,  \mbox{J.-W. Qiu}, D. Richards, R. Sufian, T. Khan     and S. Zhao for 
interest   in  our   work and  discussions. This work is supported by Jefferson Science Associates,
 LLC under  U.S. DOE Contract \#DE-AC05-06OR23177
 and by U.S. DOE Grant \#DE-FG02-97ER41028.

 \newpage 

  \appendix 
  \section{Box diagram with arbitrary indices}

 The full result for a forward matrix element is
\begin{align*}
{\mathcal O}_{\mu \alpha;  \nu \beta }^{\rm box} \to & \,
\frac{g^2N_c\Gamma(d/2)}{8\pi^2\left(-z^2\right)^{d/2}}  \int_0^1 \dd u \left(z_\mu z_\nu g_{\alpha\beta} - z_\alpha z_\nu g_{\mu\beta} - z_\mu z_\beta g_{\alpha\nu}+ z_\alpha z_\beta g_{\mu\nu} \right) \frac{2\bar u^3}{3}  G_{z\xi}(uz)  G_{z}^{\ \xi}(0) \nonumber \\
&+ \frac{g^2N_c\Gamma(d/2-1)}{8\pi^2\left(-z^2\right)^{d/2-1}}   \int_0^1 \dd u  \left\{ \vphantom{\frac{1}{2}} \left(g_{\alpha\beta} g_{\mu\nu} - g_{\mu\beta}g_{\nu\alpha} \right)\frac{2\bar u^3}{3} G_{z\xi}(uz)  G_{z }^{\ \xi}(0) \right. \nonumber \\
& \left. +\frac{\bar u^3}{3} \left(  g_{\alpha\beta} G_{z\nu}( uz) G_{z\mu}(0) -   g_{\mu\beta} G_{z\nu}( uz) G_{z\alpha}(0)-  g_{\alpha\nu} G_{z\beta}( uz) G_{z\mu}(0)+  g_{\mu\nu} G_{z\beta}( uz) G_{z\alpha}(0)\right)    \right. \nonumber \\
&  \left. +(2u\bar u+\frac{\bar u^3}{3})\left(  g_{\alpha\beta}G_{z\mu}(uz)  G_{z\nu}( 0) -   g_{\mu\beta}G_{z\alpha}(uz)  G_{z\nu}( 0) - g_{\alpha\nu} G_{z\mu}(uz)  G_{z\beta}( 0)+ g_{\mu\nu}  G_{z\alpha}(uz)  G_{z\beta}( 0)\right)    \right. \nonumber \\
 &  + \bar u^2 \left( \vphantom{\frac{1}{2}} \right.  z_\nu  G_{\alpha\beta}(uz)G_{z\mu}(0)  -  z_\nu  G_{\mu\beta}(uz)G_{z\alpha}(0) - z_\beta  G_{\alpha\nu}(uz)G_{z\mu}(0) + z_\beta G_{\mu\nu}(uz)G_{z\alpha}(0)      \nonumber \\ 
   &  -z_\mu  G_{z\nu}(uz) G_{\alpha\beta}(0) + z_\alpha  G_{z\nu}(uz) G_{\mu\beta}(0) +z_\mu  G_{z\beta}(uz) G_{\alpha\nu}(0)-z_\alpha G_{z\beta}(uz) G_{\mu\nu}(0)\left. \vphantom{\frac{1}{2}} \right)    \nonumber \\
  &  +\bar u(1+u) \left(\vphantom{\frac{1}{2}} \right.  z_\nu  G_{z\mu}(uz) G_{\alpha\beta}(0) -  z_\nu G_{z\alpha}(uz) G_{\mu\beta}(0) - z_\beta  G_{z\mu}(uz) G_{\alpha\nu}(0)+ z_\beta   G_{z\alpha}(uz) G_{\mu\nu}(0)    \nonumber \\ 
 &  -z_\mu G_{\alpha\beta}(uz)G_{z\nu}(0)  + z_\alpha  G_{\mu\beta}(uz)G_{z\nu}(0) +z_\mu   G_{\alpha\nu}(uz)G_{z\beta}(0)-z_\alpha  G_{\mu\nu}(uz)G_{z\beta}(0)\left. \vphantom{\frac{1}{2}} \right)      \nonumber \\
 &  +  (\frac{\bar u^2}{2} -\frac{\bar u^3}{3}) \left( \vphantom{\frac{1}{2} }\right. z_\nu g_{\alpha\beta} G_{\mu\xi}(uz)  G_{z }^{\ \xi}(0)  -  z_\nu g_{\mu\beta}G_{\alpha\xi}(uz)  G_{z }^{\ \xi}(0)  - z_\beta g_{\alpha\nu} G_{\mu\xi}(uz)  G_{z }^{\ \xi}(0) + z_\beta g_{\mu\nu} G_{\alpha\xi}(uz)  G_{z }^{\ \xi}(0)  \nonumber \\
  &  + z_\nu g_{\alpha\beta}G_{z\xi}(uz)  G_{\mu }^{\ \xi}(0)  -  z_\nu g_{\mu\beta} G_{z\xi}(uz)  G_{\alpha }^{\ \xi}(0) - z_\beta g_{\alpha\nu}G_{z\xi}(uz)  G_{\mu }^{\ \xi}(0) + z_\beta g_{\mu\nu} G_{z\xi}(uz)  G_{\alpha }^{\ \xi}(0)    \nonumber \\
 &   + z_\mu  g_{\alpha\beta} G_{\nu\xi}(uz)  G_{z }^{\ \xi}(0) - z_\alpha  g_{\mu\beta}G_{\nu\xi}(uz)  G_{z }^{\ \xi}(0) -z_\mu  g_{\alpha\nu} G_{\beta\xi}(uz)  G_{z }^{\ \xi}(0)+z_\alpha  g_{\mu\nu} G_{\beta\xi}(uz)  G_{z }^{\ \xi}(0)   \nonumber \\
  &   +  z_\mu g_{\alpha\beta}G_{z\xi}(uz)  G_{\nu }^{\ \xi}(0)   - z_\alpha  g_{\mu\beta}G_{z\xi}(uz)  G_{\nu }^{\ \xi}(0)  -z_\mu  g_{\alpha\nu} G_{z\xi}(uz)  G_{\beta }^{\ \xi}(0) +z_\alpha  g_{\mu\nu} G_{z\xi}(uz)  G_{\beta }^{\ \xi}(0) \left. \vphantom{\frac{1}{2} } \right)  \nonumber \\
&   + 2 \bar u \left( z_\mu z_\nu  G_{\alpha\xi}(uz) G_{\beta }^{\ \xi}(0)- z_\alpha z_\nu G_{\mu\xi}(uz) G_{\beta }^{\ \xi}(0)- z_\mu z_\beta  G_{\alpha\xi}(uz) G_{\nu }^{\ \xi}(0)+z_\alpha z_\beta  G_{\mu\xi}(uz) G_{\nu }^{\ \xi}(0) \right)    \nonumber \\
&  \left. -\frac{\bar u^3}{6}  \left( z_\mu z_\nu g_{\alpha\beta} - z_\alpha z_\nu g_{\mu\beta} - z_\mu z_\beta g_{\alpha\nu} + z_\alpha z_\beta g_{\mu\nu} \right)  G_{\zeta\xi}(uz) G^{\zeta\xi}(0) \right\} \nonumber \\
& + \frac{g^2 N_c\Gamma(d/2-2)}{8\pi^2\left(-z^2\right)^{d/2-2}} \int_0^1 \dd u   \left\{\vphantom{\frac{1}{2}} \right.    - \bar u\left(  G_{\alpha\beta}(uz)G_{\mu\nu}(0) -G_{\mu\beta}(uz)G_{\alpha\nu}(0) - G_{\alpha\nu}(uz)G_{\mu\beta}(0) + G_{\mu\nu}(uz)G_{\alpha\beta}(0)\right)    \nonumber \\
&  - 2u    G_{\mu\alpha}(uz)G_{\nu\beta}(0)   +  \bar u(1-2u)    G_{\nu\beta}(uz)  G_{\mu\alpha}(0)+\bar u(1+2u)    G_{\mu\alpha}(uz) G_{\nu\beta}(0)\nonumber \\
&  + \frac{\bar uu^2}{2}\left(  g_{\alpha\beta} G_{\mu\xi}(uz) G_{\nu }^{\ \xi}(0)   -   g_{\mu\beta}G_{\alpha\xi}(uz) G_{\nu }^{\ \xi}(0)  - g_{\alpha\nu}  G_{\mu\xi}(uz) G_{\beta }^{\ \xi}(0) +  g_{\mu\nu} G_{\alpha\xi}(uz) G_{\beta }^{\ \xi}(0)  \right)         \nonumber \\
&  + \frac{\bar uu^2}{2}\left( g_{\alpha\beta} G_{\nu\xi}(uz)  G_{\mu }^{\ \xi}(0) -   g_{\mu\beta} G_{\nu\xi}(uz)  G_{\alpha }^{\ \xi}(0)- g_{\alpha\nu}  G_{\beta\xi}(uz)  G_{\mu }^{\ \xi}(0)+  g_{\mu\nu} G_{\beta\xi}(uz)  G_{\alpha }^{\ \xi}(0) \right)         \nonumber \\
& +\bar u \left( g_{\mu\nu}  G_{\alpha\xi}(uz) G_{\beta }^{\ \xi}(0) -  g_{\alpha\nu} G_{\mu\xi}(uz) G_{\beta }^{\ \xi}(0)  - g_{\mu\beta}  G_{\alpha\xi}(uz) G_{\nu }^{\ \xi}(0) + g_{\alpha\beta} G_{\mu\xi}(uz) G_{\nu }^{\ \xi}(0)  \right) \nonumber \\
&  - \left( g_{\mu\nu} g_{\alpha\beta}   -  g_{\nu\alpha} g_{\mu\beta}   \right) \frac{\bar u^3}{6} G_{\zeta\xi}(uz) G^{\zeta\xi}(0)   \left. \vphantom{\frac{1}{2}} \right\} \ .   \hspace{8.5cm} {\rm (A.1)}
\label{boxFull}
\end{align*}

\section{Gluon-quark contribution  with arbitrary indices}

\begin{align*}
iG_{\mu\alpha}\left( z \right) G_{\nu\beta} \left( 0 \right) \to & \frac{ g^2 C_F\Gamma(d/2)z_\mu z_\nu}{8\pi^2\left( -z^2\right)^{d/2}}  \int_0^1 \dd u  \bar u  \bar\psi_c( uz)\left[   - g_{\alpha\beta} z_{\eta}  \gamma^\eta -i\epsilon_{\alpha z \beta\eta} \gamma^\eta \gamma_5\right] \psi_c(0)  \nn
& +\frac{ g^2 C_F\Gamma(d/2-1)}{16\pi^2\left( -z^2\right)^{d/2-1}}  \int_0^1 \dd u \left\{ \vphantom{1 \over } 
g_{\mu\nu} \bar u \bar\psi_c( uz) \left[  \left(z_{\alpha} g_{\beta\eta} +z_{\beta} g_{\alpha\eta} - g_{\alpha\beta} z_{\eta} \right) \gamma^\eta -i\epsilon_{\alpha z \beta\eta} \gamma^\eta \gamma_5\right] \psi_c(0) \right.   \nn
& \left. +     z_\nu\left(\vphantom{1 \over }   \bar uu \bar\psi_c( uz)\overleftarrow\partial_\mu  \left[  \left(z_{\alpha} g_{\beta\eta}  - g_{\alpha\beta} z_{\eta} \right) \gamma^\eta -i\epsilon_{\alpha z \beta\eta} \gamma^\eta \gamma_5\right] \psi_c(0) \right.  \right. \nn
&\left.  \left. + \bar u\bar\psi_c( uz)\left[ \left(g_{\mu\beta} g_{\alpha\eta} - g_{\alpha\beta} g_{\mu\eta} \right) \gamma^\eta -i\epsilon_{\alpha\mu\beta\eta} \gamma^\eta \gamma_5 \right] \psi_c(0)  \vphantom{1 \over }  \right) \right. \nn
& \left. +  z_\mu \left( \vphantom{1 \over } \bar uu \bar\psi_c (uz)\overleftarrow\partial_\nu  \left[  \left(z_{\beta} g_{\alpha\eta} - g_{\alpha\beta} z_{\eta} \right) \gamma^\eta -i\epsilon_{\alpha z \beta\eta} \gamma^\eta \gamma_5\right] \psi_c(0) \right.  \right. \nn
& \left.  \left.  +\bar u\bar\psi_c( uz) \left[ \left(g_{\alpha\nu} g_{\beta\eta}  - g_{\alpha\beta} g_{\nu\eta} \right) \gamma^\eta -i\epsilon_{\alpha\nu\beta\eta} \gamma^\eta \gamma_5 \right] \psi_c(0) \vphantom{1 \over }  \right)  \right. \nn
& \left.  -   z_\mu z_\nu \bar u^2  \left[\bar\psi_c(uz)\left(\overleftarrow\partial_\alpha \gamma_\beta + \gamma_\alpha \overleftarrow\partial_\beta \right)\psi_c(0 )  \right]  
\vphantom{1 \over } \right\}\nn
&  + \frac{g^2 C_F\Gamma(d/2-2)   }{32\pi^2\left( -z^2\right)^{d/2-2}} \int_0^1 \dd u \left\{ \vphantom {1 \over }
- g_{\mu\nu}  \bar u^2  \bar\psi_c(uz)\left(   \overleftarrow\partial_\alpha \gamma_\beta + \gamma_\alpha   \overleftarrow\partial_\beta \right)\psi_c(0)  \right. \nn
& \left. + \bar uu^2\bar\psi_c( uz)\overleftarrow\partial_\nu\overleftarrow\partial_\mu  \left[  \left(z_{\alpha} g_{\beta\eta} +z_{\beta} g_{\alpha\eta} - g_{\alpha\beta} z_{\eta} \right) \gamma^\eta -i\epsilon_{\alpha z \beta\eta} \gamma^\eta \gamma_5\right]\psi_c(0) \right. \nn
& \left. +\bar uu \bar\psi_c( uz) \overleftarrow\partial_\nu \left[ \left(g_{\mu\beta} g_{\alpha\eta} - g_{\alpha\beta} g_{\mu\eta} \right) \gamma^\eta -i\epsilon_{\alpha\mu\beta\eta} \gamma^\eta \gamma_5 \right] \psi_c(0) \right. \nn
&\left.  +\bar uu \bar\psi_c( uz)\overleftarrow\partial_\mu\left[ \left(g_{\alpha\nu} g_{\beta\eta} - g_{\alpha\beta} g_{\nu\eta} \right) \gamma^\eta -i\epsilon_{\alpha\nu\beta\eta} \gamma^\eta \gamma_5 \right]\psi_c(0)\right. \nn
& \left.  -  u \bar u^2z_\nu  \partial_\mu  \partial_\beta  \bar\psi_c(uz)  \gamma_\alpha   \psi_c(0 )  -   u \bar u^2 z_\mu   \partial_\nu   \partial_\alpha\bar\psi_c(uz)  \gamma_\beta \psi_c(0 )  
\vphantom{1 \over } \right\}  \nn
& - h.c. -\left\{ \mu \leftrightarrow \alpha \right\}-\left\{ \nu \leftrightarrow \beta \right\} + \left\{ \mu \leftrightarrow \alpha , \nu \leftrightarrow \beta \right\}
 \hspace{6.5cm} {\rm (B.1)}
\end{align*}

  \end{widetext}
  
  \bibliography{GluonLong.bib}
\bibliographystyle{apsrev4-1}

\end{document}